# Bayesian Quantum Circuit

Yuxuan Du[*][†], Tongliang Liu[†], and Dacheng Tao[†]

[†]UBTECH Sydney AI Centre, SIT, FEIT, University of Sydney

## Abstract

Parameterized quantum circuits (PQCs), as one of the most promising schemes to realize quantum machine learning algorithms on near-term quantum computers, have been designed to solve machine earning tasks with quantum advantages. In this paper, we explain why PQCs with different structures can achieve generative tasks in the discrete case from the perspective of probability estimation. Although different kinds of PQCs are proposed for generative tasks, the current methods often encounter the following three hurdles: (1) the mode contraction problem; (2) unexpected data are often generated with a high proportion; (3) target data cannot be sampled directly. For the purpose of tackling the above hurdles, we devise Bayesian quantum circuit (BQC) through introducing ancillary qubits to represent prior distributions. BQC advances both generative and semi-supervised quantum circuit learning tasks, where its effectiveness is validated by numerical simulations using the Rigetti Forest platform.

## 1 Introduction

There is a ubiquitous belief that quantum computers have the potential to bring quantum advantages to solve machine learning problems, such as achieving exponential speedups or reducing sample complexities [5]. Considering that constructing a fault-tolerant quantum computer is challenging, it is impractical to apply fault-tolerant quantum machine learning algorithms to solve realistic problems in the near future. While in the Noisy Intermediate-Scale Quantum (NISQ) era [17], it is natural to ask if the noisy quantum circuits can supply quantum advantages that attribute to tackle machine learning problems.

Recently, several studies have provided positive responses, where both the generative and discriminative tasks are accomplished by training parametrized quantum circuits (PQCs) with a classical optimization loop. PQC is composed of a set of parameterized single and controlled single qubit gates with noise. Specifically, [7, 18, 11, 9] employ PQCs with different structures to achieve classification tasks; [2, 12, 3] treat PQCs as sampling engines to generate data, in which the number of parameters can be logarithmically reduced compared with their classical counterparts.

---

[*]yudu5543@uni.sydney.edu.au

Although various PQCs are proposed, one critical issue is still remaining obscured: why PQCs can accomplish generative tasks on near-term quantum computers. A rigorous explanation is desired, which helps propose effective quantum machine learning algorithms. Thanks to the inherent nature of quantum mechanics, in this paper, we answer the above question from the perspective of probabilistic theory. Motivated by [6], we prove that existing PQCs follow the probabilistic rule for solving generative tasks in the discrete case. More specifically, for generative tasks, PQCs aim to approximate the data distribution $P(\boldsymbol{x}; \boldsymbol{\lambda})$ by updating the parameters $\boldsymbol{\theta}$ in a set of parameterized quantum gates, where $\boldsymbol{x}$ is the input data and $\boldsymbol{\lambda}$ stands for parameters.

Regarding approximating data distribution by existing PQCs, we find three deficiencies exist in generated results. For example, given training data sampled from a uniform distribution, the distribution of the generated data may be very heterogeneous. Some data are generated with a very low probability. Such a phenomenon is called mode contraction, which implies that the quantum circuit does not exactly capture the real distribution of the given data. Moreover, we find that, with the increasing dimensionality, invalidate data in the generated results may occupy a high proportion. Thirdly, existing PQCs do not support to generate the target data directly. To solve the above problems, Bayesian quantum circuit (BQC) is proposed to take the advantage of prior distributions in this paper. Specifically, instead of directly approximating the data distribution $P(\boldsymbol{x}; \boldsymbol{\lambda})$ that is adopted by most PQCs, BQC transforms to approximate the joint distribution, i.e., $P(\boldsymbol{x}, \boldsymbol{\lambda}) = P(\boldsymbol{\lambda})P(\boldsymbol{x}|\boldsymbol{\lambda})$, through introducing ancillary qubits to represent the prior distribution $P(\boldsymbol{\lambda})$. Note that $\boldsymbol{\lambda}$ can be treated as a latent variable, such as for clusters or classes. It is noteworthy that BQC makes use of a variation approach [16] to approximate the distributions $P(\boldsymbol{\lambda})$ and $P(\boldsymbol{x}|\boldsymbol{\lambda})$ by two sets of single gates and controlled single gates, respectively.

Comparing with PQCs, the proposed BQC has the following two advantages. Firstly, through validating by numerical simulations, we find that BQC circumvents the above three hurdles and outperforms state-of-the-art PQCs for specific generative tasks. Secondly, BQC can not only exploit priors to improve the performance of a learning task, but also enable us to estimate prior distributions from given data, which is essential for many learning tasks. For example, estimating class prior distribution is critical for semi-supervised learning [1]. To the best of our knowledge, BQC is the first quantum circuit scheme that can estimate prior distributions from given data. A toy model is developed to verify its effectiveness.

The rest of the paper is organized as follows: Section 2 reviews Bayesian inference by Dirac notation and PQCs; Section 3 explains why PQCs can accomplish generative tasks from the perspective of probabilistic estimation; Section 4 elaborates BQC; Section 5 demonstrates its applications and validates the proposed algorithms by experiments on numerical simulators; and Section 6 concludes the paper and discusses the future works.

## 2 Backgrounds

In classical statistics, only the observed value is considered as the variables, while $\boldsymbol{\lambda}$ are thought to be parameters. The probability density function is denoted as $P(\boldsymbol{x}; \boldsymbol{\lambda})$. In contrast, the key idea of Bayesian inference is to regard both $\boldsymbol{x}$ and $\boldsymbol{\lambda}$ as random variables. The joint probability density function is denoted as $P(\boldsymbol{\lambda}, \boldsymbol{x}) = P(\boldsymbol{\lambda})P(\boldsymbol{x}|\boldsymbol{\lambda})$, where $P(\boldsymbol{\lambda})$ and $P(\boldsymbol{x}|\boldsymbol{\lambda})$ can be used to represent a prior distribution and a likelihood function, respectively.

Regarding Bayes' rule $P(\boldsymbol{\lambda}|\boldsymbol{x}) = P(\boldsymbol{\lambda})P(\boldsymbol{x}|\boldsymbol{\lambda})/\int_{\boldsymbol{\lambda}} P(\boldsymbol{\lambda})P(\boldsymbol{x}|\boldsymbol{\lambda})d\boldsymbol{\lambda}$, the $P(\boldsymbol{\lambda}|\boldsymbol{x})$ stands for the posterior distribution of $\boldsymbol{\lambda}$ given the observed value $\boldsymbol{x}$. For the discrete case, the integral symbol in the denominator is replaced by the summation.

### 2.1 Bayesian Inference by Dirac Notation

Since quantum mechanics is intrinsically probabilistic, [6] formulates Bayesian inference by a quantum language in the discrete case. Specifically, Bayes' rule is expressed by Dirac notation [15], which is broadly employed in quantum mechanics.

Suppose the cardinality of an event space $\mathcal{S}$ equals $N$, i.e., $\mathcal{S}$ is composed with $N$ elementary events $\mathcal{S} = \{z_1, z_2, ..., z_N\}$. Each event stands for a one dimensional subspace with the unit length and is orthogonal to the rest events. Then all events form an $N$ dimensional Hilbert space. Alternatively, each event $z_i$ ($i \in [1, N]$) can be expressed by an unit basis vector, which can also be denoted as $|z_i\rangle$ by Dirac notation. Note that Dirac notation is described by a set of operators and vectors. For example, the column vector $[1, 0]^\top$ is denoted as $|0\rangle$ and the row vector $[1, 0]$ is denoted as $\langle 0|$. Using Dirac notation, the set of all events $S$ can be expressed as $|S\rangle = \sum_{z_i \in \mathcal{S}} \alpha_i |z_i\rangle$, where $|z_i\rangle$ is an $N$-dimensional unit vector, $\alpha_i$ represents the probability amplitude of the $i$-th event $z_i$. The probability of the i-th event is equal to the square of the absolute value of the $i$-th probability amplitude. Specifically, since all events are orthogonal to each other, the probability of the i-th event is $P(z_i) = \|\alpha_i |z_i\rangle\|^2 = \alpha_i^2$ with $\sum_{z_i \in \mathcal{S}} P(z_i) = \sum_{z_i \in \mathcal{S}} \|\alpha_i |z_i\rangle\|^2 = 1$.

For determining the probability of an event $A$, where $A$ is a subset of $\mathcal{S}$, a projection operator $\mathcal{P}_A$ is constructed, i.e.,

$$\mathcal{P}_A = \sum_{z_i \in A} |z_i\rangle \langle z_i| \ , \qquad (1)$$

where $\langle z_i|$ stands for the conjugate transpose of $|z_i\rangle$. The probability of an arbitrary event $A$ is defined by $P(A) = \|\mathcal{P}_A |s\rangle\|^2$, i.e.,

$$\|\mathcal{P}_A |S\rangle\|^2 = \left\| \sum_{z_i \in A} |z_i\rangle \langle z_i| \sum_{z_j \in \mathcal{S}} \alpha_j |z_j\rangle \right\|^2 = \left\| \sum_{z_i \in A} \sum_{z_j \in \mathcal{S}} \alpha_j \delta_{z_i, z_j} |z_i\rangle \right\|^2 = \sum_{z_i \in A} \alpha_i^2 = \sum_{z_i \in A} P(z_i) \ . \qquad (2)$$

Analogous to Eq. (2), the terms of Bayes' rule can be rewritten by Dirac notation. Suppose the event space $\mathcal{S}$ is of size $N$, where $N = m \cdot n$ and $m$ (or $n$) stands for that $\boldsymbol{x}$ (or $\boldsymbol{\lambda}$) can take $m$ (or $n$) different values. The prior probability distribution is then defined as $P(\boldsymbol{\lambda}) = \|\mathcal{P}(\boldsymbol{\lambda})|s\rangle\|^2$. To obtain the conditional probability $P(\boldsymbol{x}|\boldsymbol{\lambda})$, two projection operators are involved. Firstly, the quantum state $|S\rangle$ that represents the set of all events is initially interacted with the projection operator $\mathcal{P}(\boldsymbol{\lambda})$ and the quantum state transforms to $|\psi\rangle = \mathcal{P}(\boldsymbol{\lambda})|s\rangle$. Then, the second projection operator $\mathcal{P}(\boldsymbol{x})$ is employed to interact with $|\psi\rangle$. The output quantum state equals $\mathcal{P}(\boldsymbol{x})|\psi\rangle = \mathcal{P}(\boldsymbol{x})\mathcal{P}(\boldsymbol{\lambda})|s\rangle/\|\mathcal{P}(\boldsymbol{\lambda})|s\rangle\|^2$, where the conditional probability $P(\boldsymbol{x}|\boldsymbol{\lambda})$ equals $\|\mathcal{P}(\boldsymbol{x})\mathcal{P}(\boldsymbol{\lambda})|s\rangle\|^2/\|\mathcal{P}(\boldsymbol{\lambda})|s\rangle\|^2$. Likewise, the posterior equals $\|\mathcal{P}(\boldsymbol{x})\mathcal{P}(\boldsymbol{\lambda})|s\rangle\|^2/\|\mathcal{P}(\boldsymbol{x})|s\rangle\|^2$.

## 2.2 Parametric Quantum Circuits

Quantum circuit, as a model of quantum computation, is broadly employed to describe how a quantum algorithm solves a given problem [15]. In general, the key elements for a quantum circuit are classical resources, a suitable state space, ability to prepare states in the computational basis, ability to perform quantum gates, and ability to take measurements [15].

Prior to detail the parameterized quantum circuit (PQC), we introduce some notations. A fundamental element of a quantum circuit is quantum bits (qubits). Analogous to the classical computer that encodes data into bits, in quantum computing, the data is encoded into qubits. A qubit contains two basis states, $|0\rangle$ and $|1\rangle$, where $|0\rangle = [1,0]^\top$ and $|1\rangle = [0,1]^\top$ A major difference between bit and qubit is that a qubit allows a superposition of basis states, e.g., a qubit $|\psi\rangle$ can be expressed by $|\psi\rangle = \alpha_0 |0\rangle + \alpha_1 |1\rangle$ with $\|\alpha_0\|^2 + \|\alpha_1\|^2 = 1$. The state space grows exponentially with increasing the number of qubits, i.e., a system with $n$ qubit has a state space of size $2^n$. The computational basis is a product state of the basis state $\{|0\rangle, |1\rangle\}$, i.e., a computation basis for $n$ qubits is denoted as $|x_1, x_2, ..., x_n\rangle = |x_i\rangle^{\otimes n}$, where $x_i = 0, 1$, $i \in \mathbb{Z}^n$, and $\otimes$ stands for the tensor product. An $n$ qubits (quantum) gate is a basic quantum circuit operating on $n$ qubits. Mathematically, an $n$ qubit gate can be expressed as a $2^n \times 2^n$ unitary matrix. After interacting input qubits with a set of quantum gates, measurements can be performed using the computational basis of one or more of the qubits.

PQC can be regarded as a special case of the quantum circuit with a hybrid training scheme. In particular, for PQC, parts of quantum gates contain a set of (classical) parameters $\boldsymbol{\theta}$, which are updated by a hybrid quantum-classical optimization method to minimize the loss between the output and the target results. The basic framework of PQC for generative tasks is described in Supplementary Material. Employing PQC to accomplish generative learning task is as follows: (1) the number of qubits, $n$, is first determined to construct a suitable state space, where $n$ generally corresponds to the dimensions of the input data; (2) a set of quantum gates with or without parameters are employed to compose the quantum circuit with a specific structure, where the parameters $\boldsymbol{\theta}$ of parameterized gates are randomly initialized; (3) after interacting input $n$ qubits with the quantum circuit, the output quantum state is measured by the computation basis; (4) through calculating the loss between the measurement results and target results, parameters $\boldsymbol{\theta}$ are updated to minimize the loss; (5) the steps (3)-(4) are repeatedly performed until the loss is converged.

## 3 A Probabilistic Explanation for Parameterized Quantum Circuits

Although kinds of PQCs with different structures are proposed for generative tasks and are validated to be successful by empirical simulations, a fundamental question is still remained obscure: why can PQC achieve generative tasks? During composition of this manuscript, the work [11] provides an explanation from the perspective of many-body physics. However, in this paper, we build the bridge between PQCs and probabilistic theory.

For generative tasks, $n$ qubits, denoted as $|0\rangle^{\otimes n}$, are employed to represent the $2^n$-dimensional given data. In addition, the $n$ qubits form a Hilbert space with total $2^n$ orthogonal unit basis vectors. As stated in Section 2, since each event can be treated as an unit basis vector, an event space $\mathcal{S} = \{x_1, x_2, ..., x_{2^n}\}$ is constructed. In particular, the probability of the event $x_1 = |0\rangle^{\otimes n}$, is equal to one and the probability of the rest $2^n - 1$ events is equal to zero. Suppose the target probability for each event $x_i$, with $i \in \{1, 2, ..., 2^n\}$, is $P(x_i)$. Define the target probability distribution as $P(X)$. PQC can be treated as a map operator $\mathcal{P}_X$ that maps the initial probability distribution to the target distribution, i.e.,

$$\mathcal{P}_X = \sum_{x_i \in \mathcal{S}} \alpha_i |x_i\rangle \langle 0|^{\otimes n} , \qquad (3)$$

where the square of the absolute value of the $i$-th probability amplitude is equal to the target probability of the $i$-th event, with $\|\alpha_i\|^2 = P(x_i)$. It can be proven that $\mathcal{P}_X$ can map the initial distribution $P(x_1) = 1$ to the target distribution, because for each event $x_i$, the probability $P(x_i) = \| \langle x_i | \mathcal{P}_X \rangle |0\rangle^{\otimes n} \|^2 = \|\alpha_i\|^2 = P(x_i)$ and the total probability is $\left\| \mathcal{P}_X |0\rangle^{\otimes n} \right\|^2 = \left\| \sum_{x_i \in \mathcal{S}} \alpha_i |x_i\rangle \langle 0|^{\otimes n} |0\rangle^{\otimes n} \right\|^2 = \left\| \sum_{x_i \in \mathcal{S}} \alpha_i |x_i\rangle \right\|^2 = \sum_{x_i \in A} \|\alpha_i\|^2 = \sum_{x_i \in \mathcal{S}} P(x_i) = 1$.

<u>Remark 3.1.</u> The mapping operator $\mathcal{P}_X$ can be represented by a set of parameterized quantum unitary gates. Through adjusting the parameters $\boldsymbol{\theta}$, the output state of PQC can well approximate the target probability distribution $P(X)$.

For the one qubit case, with $n = 1$, the input qubit is $|0\rangle$, the cardinality of the event space $\mathcal{S}$ is 2, where $x_1 = |0\rangle$ and $x_2 = |1\rangle$, and the mapping operator $\mathcal{P}_X$ is equivalent to the Cartesian rotation gates along the $Y$-axis, denoted as $R_Y(\theta)$, specifically, $R_Y(\theta) = \begin{pmatrix} \cos(\theta/2) & -\sin(\theta/2) \\ \sin(\theta/2) & \cos(\theta/2) \end{pmatrix}$. By employing Dirac notion, we have

$$R_Y(\theta) = \cos\frac{\theta}{2}|0\rangle\langle 0| - \sin\frac{\theta}{2}|0\rangle\langle 1| + \sin\frac{\theta}{2}|1\rangle\langle 0| + \cos\frac{\theta}{2}|1\rangle\langle 1| . \qquad (4)$$

Since $\langle i|j\rangle = \delta_{i,j}$, interacting $R_Y(\theta)$ with the input qubit $|0\rangle$, we have

$$R_Y(\theta)|0\rangle = \left(\cos\frac{\theta}{2}|0\rangle\langle 0| - \sin\frac{\theta}{2}|0\rangle\langle 1| + \sin\frac{\theta}{2}|1\rangle\langle 0| + \cos\frac{\theta}{2}|1\rangle\langle 1|\right)|0\rangle$$
$$= \cos\frac{\theta}{2}|0\rangle + \sin\frac{\theta}{2}|1\rangle \ . \tag{5}$$

As shown in Eq. (5), the probability amplitudes for events $x_1 = |0\rangle$ and $x_2 = |1\rangle$ are $\cos(\theta/2)$ and $\sin(\theta/2)$, respectively. Through adjusting $\theta$, it can be guaranteed that the target distribution, $\alpha_0 = \cos(\theta/2)$ and $\alpha_1 = \sin(\theta/2)$, is approximated.

As for the multiple qubits case, with $n \geq 2$, if we prepare $n$ $R_Y(\theta)$ gates to interact with the n qubits separately, i.e., the $n$ $R_Y(\theta)$ gates are denoted as $\{R_Y(\theta_1) \otimes \mathcal{I}^{\otimes n-1}, \mathcal{I} \otimes R_Y(\theta_2) \otimes \mathcal{I}^{\otimes n-2}, ..., R_Y(\theta_n) \otimes \mathcal{I}^{\otimes n-1}\}$, many constraints are posted and the corresponding probability distribution space is very limited that may fail to approximate the target distribution $P(X)$.

For example, we start the discussion with two qubits case. When two $R_Y(\theta)$ gates, denoted as $R_Y(\theta_1)$ and $R_Y(\theta_2)$, operate with two qubits, the resulting quantum state is

$$|\phi\rangle = \cos\frac{\theta_1}{2}\cos\frac{\theta_2}{2}|00\rangle + \cos\frac{\theta_2}{2}\sin\frac{\theta_1}{2}|01\rangle + \sin\frac{\theta_2}{2}\cos\frac{\theta_1}{2}|10\rangle + \sin\frac{\theta_1}{2}\sin\frac{\theta_1}{2}|11\rangle \ . \tag{6}$$

If we use $|\phi\rangle$ to represent a probability distribution $P(\boldsymbol{x})$ with four possible events $\boldsymbol{x} = \{x_1, x_2, x_3, x_4\}$, where the probability for each event is $P(\boldsymbol{x} = x_i) = a_i$ for $i \in [1, 4]$, it is easy to see that not all probability distribution can be expressed by $|\phi\rangle$. With only two variables, the four equations: $\|\cos(\theta_1/2)\cos(\theta_2/2)\|^2 = a_1$, $\|\sin(\theta_1/2)\cos(\theta_2/2)\|^2 = a_2$, $\|\cos(\theta_1/2)\sin(\theta_2/2)\|^2 = a_3$, and $\|\sin(\theta_1/2)\sin(\theta_2/2)\|^2 = a_4$ may be ill-posed. Such constraints also hold for other multiple qubits case.

The above constraints can be weakened by introducing several blocks, at which a block is composed with $n$ $R_Y(\theta)$ gates and $k$ Controlled-NOT (CNOT) gates, with $k \leq n$. A CNOT gate involves two qubits. One is the control qubit and the other is the target qubit. If the control qubit is $|1\rangle$, the target qubit will be flipped; otherwise, if the control qubit is $|0\rangle$, the target qubit keeps unchanged.

We now discuss how to weaken the constraints for the multiple case. We still start the discussion with the two qubits case. When we employ four $R_Y(\theta)$ gates, denoted as $R_Y(\theta_1)$, $R_Y(\theta_2)$, $R_Y(\theta_3)$ and $R_Y(\theta_4)$ to operate with two qubits, where $R_Y(\theta_1)$ and $R_Y(\theta_2)$ operate with the first qubit and $R_Y(\theta_3)$ and $R_Y(\theta_4)$ operate with the second qubit, the four probability amplitudes of the new quantum state $|\phi'\rangle$ is as follows: $(\cos(\theta_1/2)\cos(\theta_2/2) - \sin(\theta_1/2)\sin(\theta_2/2))(\cos(\theta_3/2)\cos(\theta_4/2) - \sin(\theta_3/2)\sin(\theta_4/2))$ for $|00\rangle$, $(\cos(\theta_1/2)\cos(\theta_2/2) - \sin(\theta_1/2)\sin(\theta_2/2))(\cos(\theta_3/2)\sin(\theta_4/2) + \sin(\theta_3/2)\cos(\theta_4/2))$ for $|01\rangle$, $(\cos(\theta_1/2)\sin(\theta_2/2) + \sin(\theta_1/2)\cos(\theta_2/2))(\cos(\theta_3/2)\cos(\theta_4/2) - \sin(\theta_3/2)\sin(\theta_4/2))$ for $|10\rangle$, and $(\cos(\theta_1/2)\sin(\theta_2/2) + \sin(\theta_1/2)\cos(\theta_2/2))(\cos(\theta_3/2)\sin(\theta_4/2) + \sin(\theta_3/2)\cos(\theta_4/2))$ for $|11\rangle$. Compared with the quantum state $|\phi\rangle$, it is easy to see that the state $|\phi'\rangle$ can express more complex probability distributions $P(\boldsymbol{x})$ with four possible events $\boldsymbol{x} = \{x_1, x_2, x_3, x_4\}$. With continuously increasing the number of $R_Y(\theta)$ gates, the output quantum states can express any probability distribution. In addition, the CNOT gate can also contribute to strengthen the power of quantum circuit to fit a the probability distribution. Suppose the target probabilities are $P(\boldsymbol{x} = x_1) = a_1$ and $P(\boldsymbol{x} = x_4) = a_4$ with $a_1 + a_4 = 1$, where in the quantum circuit $x_1 = |00\rangle$ and $x_4 = |11\rangle$. With the help of a CNOT gate, one $R_Y(\theta)$ gate is enough to generate the quantum state corresponding to the target distribution. Specifically, when an $R_Y(\theta)$ operates the first qubits, we have $|\psi\rangle = \cos(\theta/2)|00\rangle + \sin(\theta/2)|10\rangle$. Then applying a CNOT gate on the quantum state $|\psi\rangle$, with the first qubit as controlled qubit, we have the quantum state $|\psi'\rangle = \cos(\theta/2)|00\rangle + \sin(\theta/2)|11\rangle$, which corresponds to the target distribution. It is easy to generalize above explanation to the multiple qubits case. With increasing the number of blocks, the target probability $P(X)$ can be well approximated.

It is noteworthy that, the Cartesian rotation gates along the $X$-axis and the $Z$-axis, denoted as $R_X(\theta)$ and $R_Z(\theta)$, can also be introduced to further strengthen the power of PQCs to approximate a broader probability distribution space [13]. Therefore, in general, the block of current PQCs is composed with a set of $R_X(\theta)$, $R_Y(\theta)$, $R_Z(\theta)$ and CNOT gates.

## 4 Bayesian Quantum Circuit

Directly employing PQCs to approximate given data distribution is challenging, especially when the dimensionality is large, e.g., mode contraction will occur as discussed in Section 5.1. Since PQCs follow the probabilistic rule, it is natural to ask if Bayesian inference can be introduced into PQCs to improve the performance of generative tasks [8]. Specifically, instead of approximating the given data distribution $P(\boldsymbol{x}; \boldsymbol{\lambda})$, PQC is expected to approximate the joint distribution $P(\boldsymbol{x}, \boldsymbol{\lambda})$, which can be written as

$$P(\boldsymbol{x}, \boldsymbol{\lambda}) = P(\boldsymbol{x}|\boldsymbol{\lambda})P(\boldsymbol{\lambda}) , \qquad (7)$$

where $P(\boldsymbol{\lambda})$ and $P(\boldsymbol{x}|\boldsymbol{\lambda})$ can be used to stand for a prior distribution and a likelihood function, respectively.

To approximate the joint data distribution described in Eq. (7), Bayesian quantum circuit (BQC) is proposed. Suppose the number of latent variables $\boldsymbol{\lambda}$ is $2^m$ and the dimension of given data is $2^n$. In order to approximate the prior distribution $P(\boldsymbol{\lambda})$, $m$ ancillary qubits are introduced in BQC. Additionally, different from most PQCs that employ $n$ qubits to approximate the given data distribution $P(\boldsymbol{x}; \boldsymbol{\lambda})$, BQC uses $n$ qubits to approximate the likelihood function $P(\boldsymbol{x}|\boldsymbol{\lambda})$ with the help of $m$ ancillary qubits. With the total $m + n$ qubits, the event space that describes all joint events between latent variables and observable values is of size $N = 2^{m+n}$.

The detailed framework of BQC is shown in Figure 1. With $n + m$ qubits, the initial quantum state is denoted as $|0\rangle^{\otimes(n+m)}$. The quantum circuit can be decomposed into two parts. For the first part, blocks $\{U(\boldsymbol{\gamma}^i)\}_{i=1}^{J}$ are employed to operate with $m$ ancillary qubits The output quantum state $|\boldsymbol{\lambda}\rangle$ corresponds to the prior distribution $P(\boldsymbol{\lambda})$. Note that $\boldsymbol{\gamma}$ represents the parameters of a set of

$R_Y(\boldsymbol{\gamma})$ gates. As for the second part, supporting by the following Remark 4.1 and Remark 4.2, the new blocks $\{U(\boldsymbol{\theta}^i)\}_{i=1}^{L}$, where each block is composed of $n$ Controlled Cartesian rotation gate along the Y axis (CRY($\theta$)) and Toffoli gates (T-gate) for each block, are constructed to approximate the likelihood function $P(\boldsymbol{x}|\boldsymbol{\lambda})$.

Specifically, CRY($\theta$) represents that: if the control qubit is $|1\rangle$, the $R_Y(\theta)$ gate is applied on the target qubit; otherwise, the target qubit keeps unchanged. The Toffoli gate is defined as: if the two control qubits are both $|1\rangle$, the target qubit will be flipped; otherwise, the target qubit keeps unchanged.

In BQC, the control qubit of CRY($\theta$) is always designed to be one of the $m$ ancillary qubits and its target qubit is always designed to be one of the $n$ data qubits. Similarly, one control qubit in T-gate must belong to $m$ ancillary qubits and the target qubit must belong to data qubits. Such a operating method constructs the connection between data $\boldsymbol{x}$ and latent variables $\boldsymbol{\lambda}$. The output quantum state of $n+m$ data qubits corresponds to the joint distribution $P(\boldsymbol{x},\boldsymbol{\lambda})$. Meanwhile, conditionally seeing ancillary qubits, the output quantum state of $n$ data qubits corresponds to $P(\boldsymbol{x}|\boldsymbol{\lambda})$. The output quantum state of $m$ ancillary qubits corresponds to $P(\boldsymbol{\lambda})$.

<u>Remark 4.1.</u> Suppose the quantum state after interacting $|0\rangle^{\otimes(n+m)}$ with the mapping operator $\mathcal{P}_{\boldsymbol{\lambda}}$ is $|\Phi\rangle$, where $\mathcal{P}_{\boldsymbol{\lambda}} = \sum_{\lambda_i \in \boldsymbol{\lambda}} \beta_i |\lambda_i\rangle \langle 0|^{\otimes n}$ and $\|\beta_i\|^2 = P(\boldsymbol{\lambda} = \lambda_i)$. Applying the mapping operator $\mathcal{P}_{\boldsymbol{x}}$ onto the quantum state $|\Phi\rangle$, where $\mathcal{P}_{\boldsymbol{x}} = \sum_{x_j \in \mathcal{S}} \sum_{\lambda_i \in \boldsymbol{\lambda}} \alpha_{ji} |x_j\rangle |\lambda_i\rangle \langle 0|^{\otimes n} \langle \lambda_i|$, the joint distribution $P(\boldsymbol{x},\boldsymbol{\lambda})$ is determined by $\|\mathcal{P}_{\boldsymbol{x}}|\Phi\rangle\|^2$. Then, the probability of a new event with $\boldsymbol{x} = x_j$ and $\boldsymbol{\lambda} = \lambda_i$ is determined by $P(\boldsymbol{x} = x_j, \boldsymbol{\lambda} = \lambda_i) = \|\alpha_{ji}\beta_i\|^2$.

<u>Remark 4.2.</u> The projection operator $\mathcal{P}_{\boldsymbol{x}}$ can be represented by a set of controlled Cartesian rotation gate along Y axis (CRY($\theta$)) and Toffoli gate (T-gate). Meantime, The CRY($\theta$) and Toffoli gate can be represented by a set of single qubit gates and controlled single qubit gates.

We discuss the above two Remarks in the Supplementary Material.

Analogous to PQCs, the loss is calculated by comparing the measurement results with given data, which is accomplished by a classical machine. Through repeatedly updating the parameters of the parameterized gates, the loss between the generated and target results is minimized. It is noteworthy that two sets of parameters $\boldsymbol{\gamma}$ and $\boldsymbol{\theta}$ are independent with each other. If the prior distribution is known, the parameters $\boldsymbol{\gamma}$ keep fixed and $\boldsymbol{\theta}$ are updated to minimize the loss in the learning process; otherwise, if the prior distribution is unknown but the likelihood is given, $\boldsymbol{\gamma}$ will be updated in the learning process. The selection of the loss function and optimization method is flexible. In this paper, the squared maximum mean discrepancy (MMD) loss and a gradient optimization method [12] are employed to update the parameters. The definition of the MMD loss and the proposed BQC algorithm are summarized in Supplementary Material.

The quantum advantage provided by BQC has two perspectives: (1) similar to PQCs, the parameters required in the proposed algorithm is exponentially reduced than its classical counterparts. Specifically, the number of parameters in blocks $\{U_i(\boldsymbol{\gamma})\}_{i=1}^{J}$ only relates to the number of latent variables, which are generally small and irrelevant to the dimension of data. Additionally, since the $n$-dimensional data can be recorded by $\log n$ qubits with amplitude encoded methods [10], the number of parameters in $L$ blocks $\{U(\boldsymbol{\theta_i})\}_{i=1}^{L}$ is polynomial logarithmic in terms of the input data dimension. Therefore, the total parameters is $O(poly(M + \log N))$, where $M$ is the number of latent variables, $M \ll N$, and $N$ is the dimension of data. (2) Different with PQCs, the prior distribution and the likelihood function can be explicitly extracted in BQC.

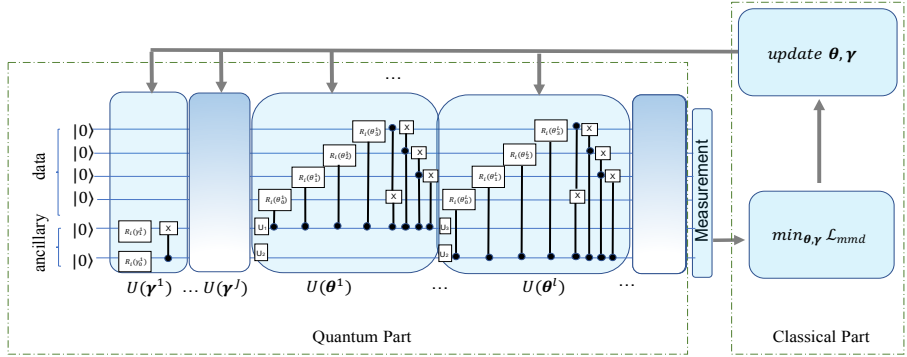

Figure 1: The proposed BQC. Note that $J$ blocks $\{U(\boldsymbol{\gamma}^j)\}_{j=1}^{J}$ are employed to generate quantum state that corresponds to the $P(\boldsymbol{\lambda})$ and $L$ blocks $\{U(\boldsymbol{\theta}^i)\}_{i=1}^{L}$ are employed to generate the quantum state that corresponds to $P(\boldsymbol{x}|\boldsymbol{\lambda})$. the blocks $\{U(\boldsymbol{\theta}^i)\}_{i=1}^{L}$ access to operate with data qubits is controlled by the ancillary qubits.

## 5 Applications

To demonstrate the advantages of the proposed BQC, two applications are discussed here. In Subsection 5.1, BQC is employed to accomplish generative tasks, e.g., to learn $P(\boldsymbol{x}|\boldsymbol{\lambda})$; in Subsection 5.2, BQC is employed to learn prior distributions from given data, i.e., to learn $P(\boldsymbol{\lambda})$. Both the generative and learning prior tasks are validated by numerical simulations. The BQC is implemented in Python, leveraging the pyQuil library to access the numerical simulator [19].

### 5.1 Generative Model Exploiting Priors

Analogous to the studies [12, 2], the bars and stripes (BAS) dataset [14] is employed as a benchmark, where the dataset is composed with vertical bars and horizontal stripes. For $n \times m$ pixels, the number of images that belongs to BAS is $N_{BAS} = 2^n + 2^m - 2$. Some BAS images of $2 \times 2$ pixels are shown in Figure 2 (d). In accordance with the conventions in previous study, in BQC, all BAS patterns are encoded in the quantum state, where each qubit stands for a pixel of the BAS image.

Following Eq. (7), a prior distribution is first encoded into the ancillary qubits. Specifically, since each BAS image is expected to be generated uniformly, the prior probability that represents each BAS image is $1/N_{BAS}$, where the number of latent variables $\boldsymbol{\lambda}$ equals $N_{BAS}$ and $m = \lceil \log N_{BAS} \rceil$ ancillary qubits are required. It is noteworthy that, in the BAS case, since each $\lambda_i$ corresponds to a specific BAS image, we have $P(\boldsymbol{x} = x_1, \boldsymbol{\lambda} = \lambda_1) = P(x_1)$. This is a special case of $P(\boldsymbol{x}|\boldsymbol{\lambda})$ for generative task.

We first train BQC on the $2 \times 2$ BAS dataset, where $N_{BAS} = 6$ valid images are expected to be generated uniformly after learning. In the experiment, the numbers of qubits to record BAS images and to represent the prior distribution are $n = 4$ and $m = 3$, respectively. Since the prior distribution is known, the parameters of blocks $\{U_j(\boldsymbol{\gamma})\}_{j=1}^{2}$ are fixed, such that $P(\boldsymbol{\lambda} = \lambda_i) = 1/\sqrt{N_{BAS}}$ for $i \in [1, 6]$. Each quantum state $|\lambda_i\rangle$, as a control state, interacts with the blocks $\{U_i(\boldsymbol{\theta})\}_{i=1}^{2}$. Total 48 parameters of $\boldsymbol{\theta}$ are updated in the learning process. Moreover, BQC is applied to generate a $3 \times 3$ BAS dataset, with $N_{BAS} = 14$. The numbers of data qubits $n$ and ancillary qubits $m$ are set to be 9 and 4, respectively. Total 112 parameters of $\boldsymbol{\theta}$ are updated in the learning process.

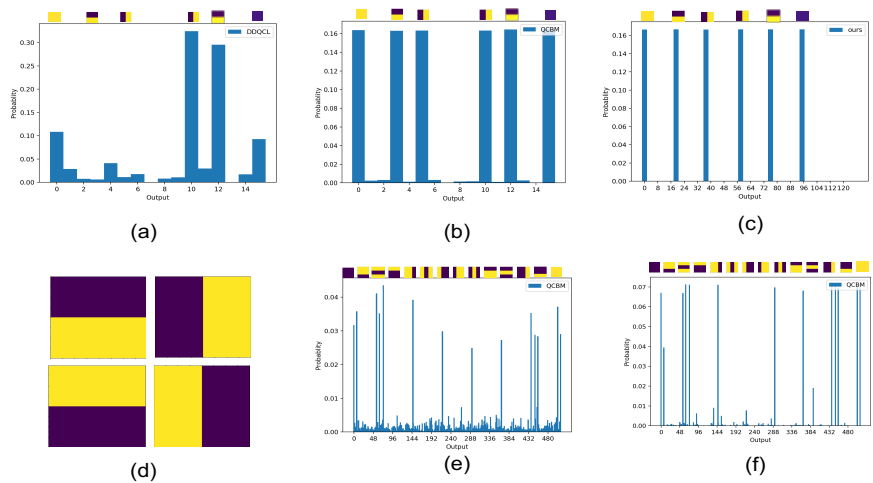

Figure 2: The generative results obtained from DDQCL, QCBM, and our model. Since the BAS dataset can be regard as a set of binary images, it can be mapped into different integers, as the x-axis of figures.

We compare the generative performance of BQC with another two PQCs, i.e., data driven quantum circuit learning (DDQCL) [2] and quantum circuit born machine (QCBM) [12]. The numbers of updated parameters are restricted into the same level. Additionally, benefiting from that the numerical simulator Rigetti Quantum Virtual Machine (QVM) [19] supports readout of the probability amplitudes, the number of measurements is set to be $\infty$, which implies the distribution of BAS images can be accessed directly. As shown in Figures 2 (a), (b), (c): for the 2 × 2 BAS dataset, after learning, there is an obvious difference between the generated and the expected distribution learned by DDQCL; while QCBM and BQC generate the expected BAS images uniformly and accurately. We further compare the generated results of 3 × 3 BAS images between QCBM and BQC. As shown in Figures 2 (e) and (f), the generated distribution of BQC is very close to the expected one, where the probability of most valid images is near to $1/N_{BAS} \approx 0.0714$. In contrast, the generated distribution of QCBM is obviously different from the expected one. The accuracies of generative valid BAS images from the three PQCs are listed in the Supplementary Material, where our model obtains the highest accuracy.

### 5.2 Learning Unknown Prior Distributions

To validate the effectiveness of BQC for learning prior distributions from given data, we devise a toy model. In the model, the training data are sampled form a joint distribution $P(\boldsymbol{x}, \boldsymbol{\lambda})$, i.e., $P(\boldsymbol{x}, \boldsymbol{\lambda}) = P(\lambda_1)\mathcal{N}_1(\mu_1, \sigma_1) + P(\lambda_2)\mathcal{N}_2(\mu_2, \sigma_2)$. Specifically, the likelihood $P(\boldsymbol{x}|\boldsymbol{\lambda})$ is known, which is denoted as $P(\boldsymbol{x}|\boldsymbol{\lambda} = \lambda_1) \sim \mathcal{N}_1(\mu_1, \sigma_1)$ and $P(\boldsymbol{x}|\boldsymbol{\lambda} = \lambda_2) \sim \mathcal{N}_2(\mu_2, \sigma_2)$, where $\mathcal{N}_1(\mu_1, \sigma_1)$ and $\mathcal{N}_2(\mu_2, \sigma_2)$ stand for two Gaussian distributions with means $\mu_1$ and $\mu_2$, variations $\sigma_1$ and $\sigma_2$, respectively. Note that $x$ in $\mathcal{N}_1$ and $\mathcal{N}_2$ is an integer variable encoded by qubits, where $x_{max} = 2^n$ and $n$ is the number of qubits. Given the sampled data and the likelihood functions, BQC is employed to estimate the prior distribution $P(\boldsymbol{\lambda})$ for the toy model.

In the learning process, since $\mathcal{N}_1(\mu_1, \sigma_1)$ and $\mathcal{N}_2(\mu_2, \sigma_2)$ are given, the parameters $\boldsymbol{\theta}$ of blocks $U(\boldsymbol{\theta})$ keep fixed. Meanwhile, since the number of $\boldsymbol{\lambda}$ is two, one ancillary is used to represent the prior distribution. The number of parameter $\boldsymbol{\gamma}$ is set to be one and $\gamma$ is updated to minimize the MMD loss.

The numerical simulation results are listed in Table 1. The number of qubits used to represent two Gaussian distributions is $n = 7$. The means and variances of the two Gaussian distributions $\mathcal{N}_1$ and $\mathcal{N}_2$ set as $u_1 = 16$, $\mu_2 = 64$, $\sigma_1 = 2$, $\sigma_2 = 4$, respectively. We estimate two sets of the target coefficients, i.e., $P(\lambda_1) = 0.7$, $P(\lambda_2) = 0.3$ and $P(\lambda_1) = 0.85$, $P(\lambda_2) = 0.15$, respectively. The number of parameters $\boldsymbol{\theta}$ to describe the two Gaussian distributions is 56. The number of measurements is set to be 200, 1000, and $\infty$, respectively. Each experiment is conducted 6 times, where the average $P(\lambda_1)$, $P(\lambda_2)$, and their variances are calculated. Since $P(\lambda_1)$ and $P(\lambda_2)$ share the same variance, we only list one in the table. The learned prior distributions are very close to the target one. The small difference is mainly caused by that the limited parameters $\boldsymbol{\theta}$ cannot approximate $\mathcal{N}_1$ and $\mathcal{N}_2$ well.

## 6  Conclusion

In this paper, we explain why PQCs with different structures can accomplish generative machine learning tasks from the perspective of probability estimation. Then, motivated by Bayesian inference, BQC is proposed. The parameters of BQC is polynomial logarithmic in terms of the dimensions of data, which is dramatically reduced than its corresponding classical machine learning algorithms. Furthermore, comparing with PQCs, BQC can extract prior distributions and likelihood function explicitly. Employing BQC to exploit and learn prior distribution of given data is also demonstrated. In the future, we plan to apply BQC to solve more complicated statistical learning problems with quantum advantages.

Table 1: Learning Prior Distribution with $m = 1$, $n = 7$

| Methods | # of Mea | $P(\lambda_1)$ | $P(\lambda_2)$ | Variance | $P(\lambda_1)$ | $P(\lambda_2)$ | Variance |
|---|---|---|---|---|---|---|---|
| Target | – | 0.70 | 0.30 | – | 0.85 | 0.15 | – |
| QVM | 200 | 0.74 | 0.26 | 2.63E-02 | 0.79 | 0.21 | 5.10E-02 |
| QVM | 1000 | 0.64 | 0.36 | 1.10E-02 | 0.80 | 0.20 | 6.95E-03 |
| QVM | $\infty$ | 0.68 | 0.32 | 0 | 0.87 | 0.13 | 0 |

# 7 Supplementary Materials

In this supplementary material, Section 8 introduces more details about general PQCs; Section 9 discuss the Remark 4.1 and Remark 4.2; Section 10 demonstrates the MMD loss and summarizes the algorithm of BQC; Section 11 illustrates the accuracies of generating the BAS dataset by different methods.

# 8 Details about PQCs

The framework of employing PQCs to accomplish learning tasks is illustrated in Figure 3. When kinds of PQCs are employed to accomplish the generative tasks, the current main differences of them are the structures of quantum circuits and the optimization methods.

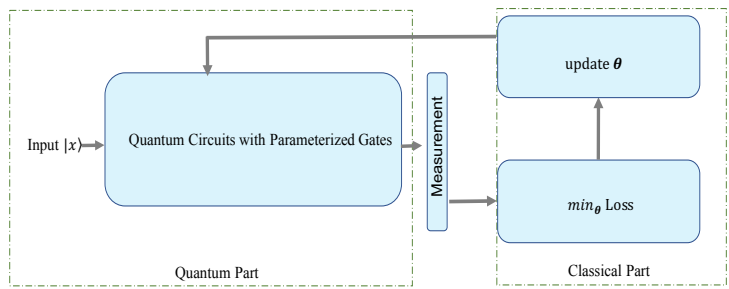

Figure 3: The parameterized quantum circuits with a classical optimization loop.

# 9 Discussion of Remark 4.1 and Remark 4.2

In a nutshell, analogous to the proof of Remark 3.1, the joint distribution $P(\boldsymbol{x}, \boldsymbol{\lambda})$ can also be formulated by Dirac notation, where two mapping operators are employed. Meanwhile, the two mapping operators can be represented by a set of single qubit gates and controlled single qubit gates, which implies the joint distribution $P(\boldsymbol{x}, \boldsymbol{\lambda})$ can be approximate by quantum circuits.

*Remark 4.1.* Analogous to the proof of Theorem 3.1, given the data with the dimension of $2^n$. Note that $n$ data qubits are involved to guarantee the sample space $\mathcal{S}$ is $2^n$ and all events are orthogonal to each other. In addition, $m$ ancillary qubits are introduced to form the latent variable space $\boldsymbol{\Lambda}$ with size $2^m$, where all possible events of $\boldsymbol{\lambda}$ are orthogonal to each other. The $n+m$ qubits guarantee that the joint events space, $\mathcal{S} \wedge \boldsymbol{\Lambda}$, is of size $2^{(n+m)}$.

When the mapping operator $\mathcal{P}_{\boldsymbol{\lambda}}$ is operated with the input $m$ ancillary qubits $|0\rangle^{\otimes m}$, the output quantum state $|\Phi\rangle$ is

$$|\Phi\rangle = \mathcal{I}^{\otimes n} \sum_{\lambda_i \in \boldsymbol{\Lambda}} \beta_i |\lambda_i\rangle \langle 0|^{\otimes m} |0\rangle^{\otimes (n+m)} = \sum_{\lambda_i \in \boldsymbol{\Lambda}} \beta_i |0\rangle^{\otimes n} |\lambda_i\rangle , \qquad (8)$$

which corresponds to the prior distribution $P(\boldsymbol{\lambda})$. Specifically, we have $\|\mathcal{P}_{\boldsymbol{\lambda}} |0\rangle^{\otimes (n+m)}\|^2 = \|\sum_{\lambda_i \in \boldsymbol{\Lambda}} \beta_i |0\rangle^{\otimes n} |\lambda_i\rangle\|^2 = \sum_{\lambda_i \in \boldsymbol{\Lambda}} \|\beta_i\|^2$, where the square of the absolute value of the $i$-th probability amplitude $\|\beta_i\|^2$ is equal to the probability $P(\lambda_i)$. Therefore, after interacting mapping operator $\mathcal{P}_{\boldsymbol{\lambda}}$ with $m$ ancillary qubits, the initial probability distribution of the $m$ ancillary qubits, i.e., $P(\boldsymbol{\lambda} = \lambda_1) = 1$ with $\lambda_1 = |0\rangle^{\otimes m}$, is mapped to the prior distribution $P(\boldsymbol{\lambda})$.

Then, the mapping operator $\mathcal{P}_{\boldsymbol{x}}$ is applied onto the quantum state $|\Phi\rangle$, the output quantum state $|\Psi\rangle$ is

$$|\Psi\rangle = \sum_{x_j \in \mathcal{S}} \sum_{\lambda_i \in \boldsymbol{\Lambda}} \alpha_{ji} |x_j\rangle |\lambda_i\rangle \langle 0|^{\otimes n} \langle \lambda_i| \sum_{\lambda_k \in \boldsymbol{\Lambda}} \beta_k |0\rangle^{\otimes n} |\lambda_k\rangle = \sum_{x_j \in \mathcal{S}} \sum_{\lambda_i \in \boldsymbol{\Lambda}} \alpha_{ji} \beta_i |x_j\rangle |\lambda_i\rangle ,$$
$$(9)$$

where the square of the absolute probability amplitude $\alpha_{ji}$, denoted as $\|\alpha_{ji}\|^2$, is equal to the conditional probability of the event $\boldsymbol{x} = x_j$ given $\boldsymbol{\lambda} = \lambda_i$, denoted as $P(\boldsymbol{x} = x_j | \boldsymbol{\lambda} = \lambda_i)$.

Eq. (9) corresponds to the joint distribution $P(\boldsymbol{x}|\boldsymbol{\lambda})$. Specifically, for any joint event $x_i \wedge \lambda_j$, its corresponding probability is $\|\langle x_i, \lambda_j|\Psi\rangle\|^2 = \|\beta_i\|^2 \|\alpha_{ji}\|^2$, which is equivalent to $P(\boldsymbol{\lambda} = \lambda_i) P(\boldsymbol{x} = x_j | \boldsymbol{\lambda} = \lambda_i)$. Therefore, the two mapping operators map the initial probability distribution $P(\boldsymbol{x} = |0\rangle^{\otimes n}, \boldsymbol{\lambda} = |0\rangle^{\otimes m}) = 1$ to the target joint distribution $P(\boldsymbol{x}, \boldsymbol{\lambda})$.

*Remark 4.2.* The quantum state $\sum_{x_j \in \mathcal{S}} \sum_{\lambda_i \in \boldsymbol{\Lambda}} \alpha_{ji} \beta_i |x_j\rangle |\lambda_i\rangle$ in Eq. 9 indicates that, given $\boldsymbol{\lambda} = \lambda_i$, the corresponding $\boldsymbol{x} = x_j$ has the specific probability amplitude $\alpha_{ji}$. In BQC, as stated in the proof of Remark 3.1, the mapping operator $\mathcal{P}_{\boldsymbol{\lambda}}$ can be constructed by a set of $R_Y(\gamma)$ gates and CNOT gates. After operating $\mathcal{P}_{\boldsymbol{\lambda}}$ with the $m$ ancillary qubits, the output quantum state of ancillary qubits corresponds to the prior distribution $P(\boldsymbol{\lambda})$. In addition, since the probability amplitudes $\alpha_{ji}$ has a strong relation with $\lambda_i$, the mapping operator $\mathcal{P}_{\boldsymbol{x}} = \sum_{x_j \in \mathcal{S}} \sum_{\lambda_i \in \boldsymbol{\lambda}} \alpha_{ji} |x_j\rangle |\lambda_i\rangle \langle 0|^{\otimes n} \langle \lambda_i|$ are constructed by employing

the controlled rotation gates along the Y-axis(CRY($\theta$)) and Toffoli gates. The control qubit of CRY($\theta$) and one control qubit of T-gate must be designed as ancillary qubits to guarantee that $\alpha_{ji}$ can be differed with various $|\lambda_i\rangle$.

Since CRY($\theta$) = $\begin{pmatrix} 1 & 0 & 0 & 0 \\ 0 & 1 & 0 & 0 \\ 0 & 0 & \cos(\theta/2) & -\sin(\theta/2) \\ 0 & 0 & \sin(\theta/2) & cos(\theta/2) \end{pmatrix}$. By Dirac notation, we have

$$\text{CRY}(\theta) = |00\rangle\langle 00| + |01\rangle\langle 01| + \cos\frac{\theta}{2}|10\rangle\langle 10| + \sin\frac{\theta}{2}|10\rangle\langle 11| + \sin\frac{\theta}{2}|11\rangle\langle 10| + \cos\frac{\theta}{2}|11\rangle\langle 11|. \tag{10}$$

Eq. 10 indicates that if and only if the first qubit equals $|1\rangle$, the $R_Y(\theta)$ gates will operate with the second qubit. When the cardinality of both sample space and latent variable space is 2, analogous to the proof of Remark 3.1, it is easy to prove the mapping operator $\mathcal{P}_{\boldsymbol{x}}$ can be expressed by two CRY($\theta$) gates and one NOT gate.

As for the high dimensional case, if $m \times n$ CRY($\theta$) gates are directly applied to interact with the $n$ data qubits, the $m \times n$ constraints make the probability space is very limited as state in the proof of Remark 3.1. Similar to the proof of Remark 3.1, CNOT gates are expected to weaken such constraints. Meanwhile, the controlled qubits $|\lambda\rangle$ is also required. Therefore, a Toffoli gate is introduced, which has two control qubits. As stated in the proof of Remark 3.1, with a set of CRY($\theta$) gates and Toffoli gates, the constraints will be weakened.

Followed the theorem of Single qubit and CNOT gates are universal [15], the CRY($\theta$) and Toffoli gate can be represented by three Cartesian rotation gates and CNOT gates.

## 10 MMD Loss and BQC

In this section, we briefly introduce the MMD loss that is employed in the paper and summarize the algorithm of the proposed BQC.

### 10.1 MMD Loss

The selection of the loss function and the optimization method is flexible in BQC. In the paper, we employ MMD as the loss function. Mathematically, suppose the parameters are $\boldsymbol{\theta}$ to update in blocks $U(\boldsymbol{\theta})$ and the parameters $\boldsymbol{\gamma}$ keeps unchanged, the MMD loss, $\mathcal{L}$, is denoted as

$$\mathcal{L} = \left\| \sum_{x^i \in \text{BAS}} P(x^i|\boldsymbol{\lambda})\phi(x^i) - \sum_{x^i \in \text{BAS}} f(x^i)\phi(x^i) \right\|^2, \tag{11}$$

where $\phi(x^i)$ maps the $i$-th input data, $x^i$, into a high-dimensional reproducing kernel Hilbert space [4], $f(x^i)$ is the expected distribution of given data, and $P(x^i|\boldsymbol{\lambda})$ is determined by measuring the first $n$ qubits of the output quantum states $|\psi\rangle$ by the computational basis $|x^i\rangle$, i.e., $P(x^i|\boldsymbol{\lambda}) = \|\langle x^i|\psi\rangle\|^2$. More details about the MMD loss and how to optimize it by employing the gradient method are introduced in [12].

### 10.2 BQC

To gain a better understanding of the proposed BQC, we summarize it in the following Algorithm 1.

---
**Algorithm 1:** Bayesian Quantum Circuits

**Input** : The given data $\boldsymbol{x}$;
           The number of data qubits and ancillary qubits, $m$ and $n$;
           Number of $J$ Blocks $\{U(\boldsymbol{\gamma}^i)\}_{i=1}^J$ and $L$ Blocks $\{U(\boldsymbol{\theta}^i)\}_{i=1}^L$;
           Tolerance $\epsilon$ ;
           Number of measurements $N$.
**Output:** The generated data or prior distributions.

1. Initializing parameters of all blocks randomly, $U(\boldsymbol{\theta})$ or $U(\boldsymbol{\gamma})$, randomly;
2. **while** $Loss > \epsilon$ **do**
3.     Applying $J$ blocks $\{U(\boldsymbol{\gamma}^i)\}_{i=1}^J$ onto the $m$ ancillary qubits and $L$ blocks $\{U(\boldsymbol{\theta}^i)\}_{i=1}^L$ onto the $n$ data qubits ;
4.     Measuring $N$ copies of the outputs;
5.     Calculating the loss between the given data and the measurement results;
6.     Updating the parameters to minimize the loss;
7. **end**

---

## 11 Accuracies on the BAS Dataset

When three PQCs, e.g., DDQCL, QCBM, and BQC (proposed), with the same parameter level, are employed to generate valid BAS images with pixels 2 and $3 \times 3$, the corresponding accuracies are listed in Table 2. As we can see, when the number of learning parameters is at the same level, BQC outperforms state-of-the-art PQCS in generating BAS datasets. The results indicates that, BQC has the least proportion that the generated images do not belong to the BAS dataset than other two methods.

Table 2: Accuracies for generative 2×2 and $3 \times 3$ BAS datasets

|  | Model | DDQCL | QCBM | BQC |
|---|---|---|---|---|
| $2 \times 2$ | Accuracy (%) | 83.82 | 98.46 | 99.96 |
| $3 \times 3$ | Accuracy (%) | – | 46.76 | 89.91 |